\def\tipo{1}

\def\av#1{\langle#1\rangle}          
\newcommand{\omitit}[1]{}

\ifnum \tipo = 2
\documentclass[twocolumn,showpacs,preprintnumbers,amsmath,amssymb,superscriptaddress]{revtex4}             

\def\subfigsize{4.2cm}
\else 
\documentclass[preprint,showpacs,preprintnumbers,amsmath,amssymb,superscriptaddress]{revtex4}

\def\subfigsize{6cm}
\fi

\usepackage{graphicx}
\usepackage{dcolumn}
\usepackage{bm}
\usepackage{subfigure}

\begin{document}

\preprint{APS/123-QED}

\title{Scaling of Optimal Path Lengths Distribution in Complex Networks}

\author{Tomer Kalisky}
\email{kaliskt@mail.biu.ac.il}
\affiliation{Minerva Center and Department of Physics, Bar-Ilan
  University, 52900 Ramat-Gan, Israel}
%
%
\author{Lidia A. Braunstein}
\affiliation{Center for Polymer Studies and Department of Physics
  Boston University, Boston, MA 02215, USA}
\affiliation{Departamento de F\'{\i}sica, Facultad de Ciencias Exactas
  y Naturales, Universidad Nacional de Mar del Plata, Funes 3350,
  $7600$ Mar del Plata, Argentina}
\author{Sergey V. Buldyrev}
\affiliation{Center for Polymer Studies
  and Department of Physics Boston University, Boston, MA 02215, USA}
\author{Shlomo Havlin}
\affiliation{Minerva Center and Department of
  Physics, Bar-Ilan University, 52900 Ramat-Gan, Israel}
\affiliation{Center for Polymer Studies and Department of Physics
  Boston University, Boston, MA 02215, USA}
\author{H. Eugene Stanley}
\affiliation{Center for Polymer Studies and Department of Physics
  Boston University, Boston, MA 02215, USA}


\date{\today} 

\begin{abstract}
  We study the distribution of optimal path lengths in random graphs
  with random weights associated with each link (``disorder''). With
  each link $i$ we associate a weight $\tau_i = \exp(ar_i)$ where
  $r_i$ is a random number taken from a uniform distribution between 0
  and 1, and the parameter $a$ controls the strength of the disorder.
  We suggest, in analogy with the average length of the optimal path,
  that the distribution of optimal path lengths has a universal form
  which is controlled by the expression
  $\frac{1}{p_c}\frac{\ell_{\infty}}{a}$, where $\ell_{\infty}$ is the
  optimal path length in strong disorder ($a \rightarrow \infty$) and
  $p_c$ is the percolation threshold. This relation is supported by
  numerical simulations for Erd\H{o}s-R\'enyi and scale-free graphs.
  We explain this phenomenon by showing explicitly the transition
  between strong disorder and weak disorder at different length scales
  in a single network.
  
\end{abstract}



\maketitle


\def\figureERWeak{
  \begin{figure}
    \resizebox{\subfigsize}{!}{\includegraphics{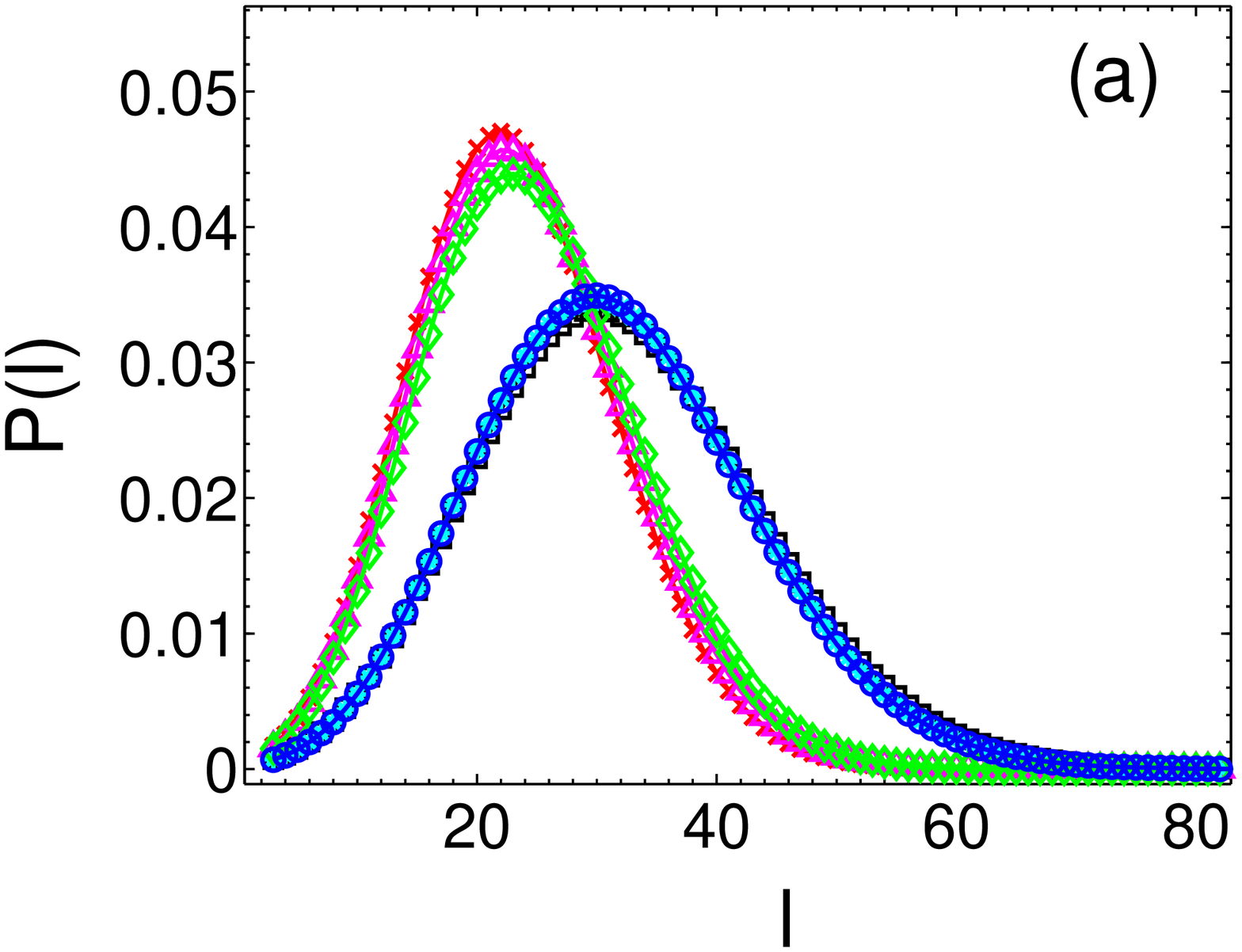}}
    \hskip 1 mm
    \resizebox{\subfigsize}{!}{\includegraphics{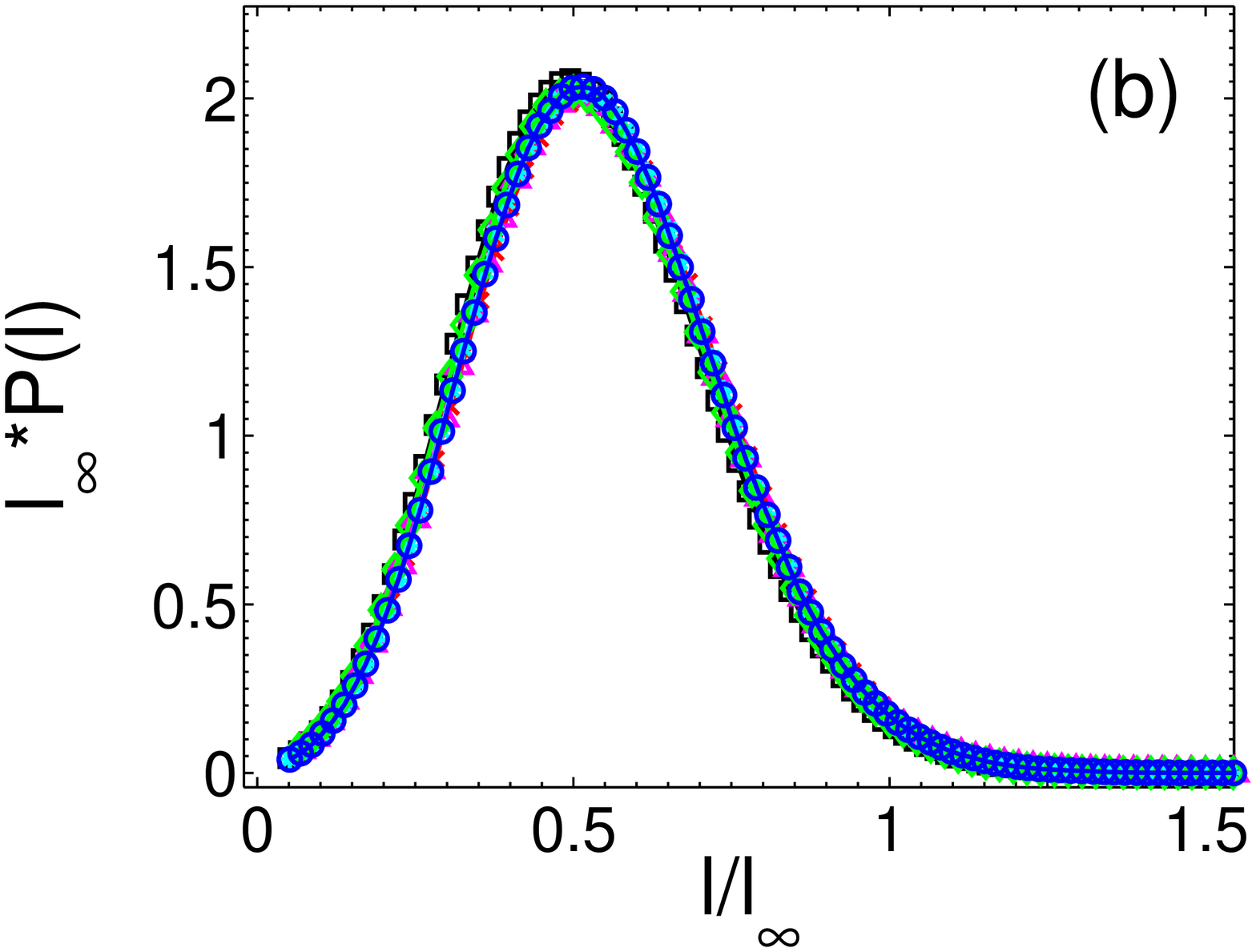}}
    \\
    \resizebox{\subfigsize}{!}{\includegraphics{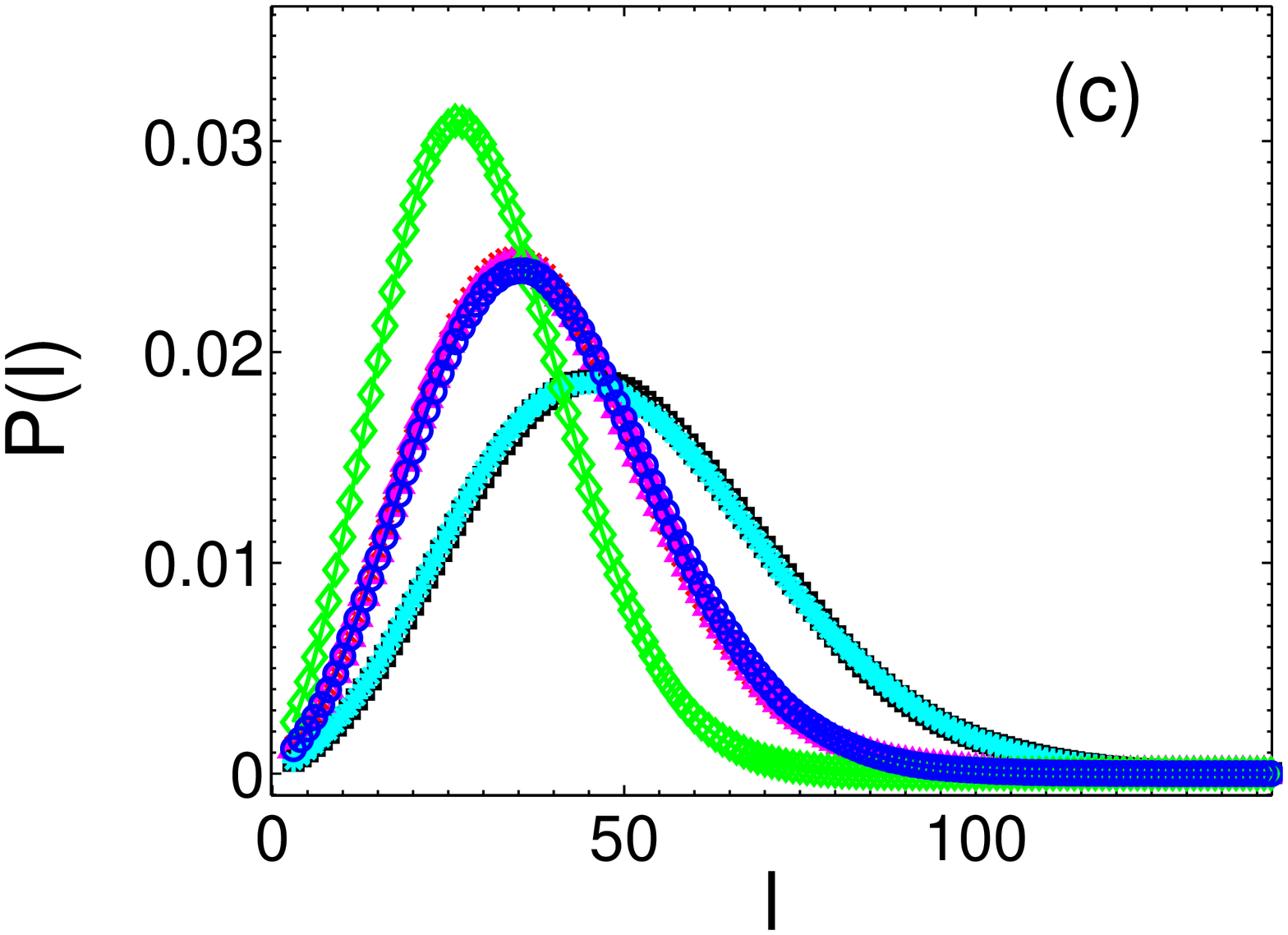}}
    \hskip 1 mm
    \resizebox{\subfigsize}{!}{\includegraphics{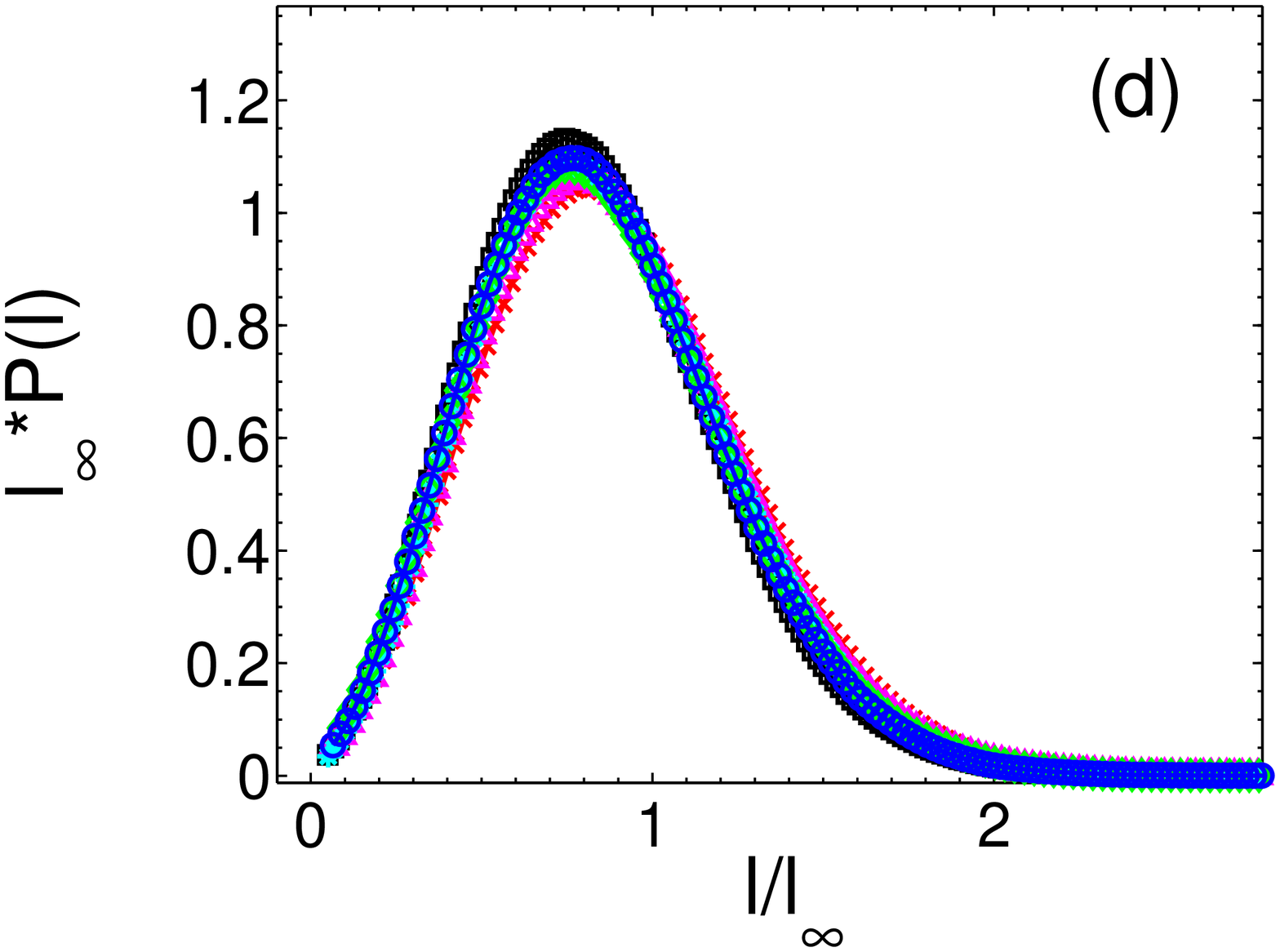}}
    \caption{\label{fig:ERWeak} (Color online) Optimal path lengths
      distribution, $P(l)$, for ER networks with (a,b) $Z \equiv
      \frac{1}{p_c}\frac{\ell_{\infty}}{a} = 10$ and (c,d) $Z =3$. (a)
      and (c) represent the unscaled distributions for $Z=10$ and
      $Z=3$ respectively, while (b) and (d) are the scaled
      distribution. Different symbols represent networks with
      different characteristics such as size $N$ (which determines
      $\ell_{\infty} \sim N^{1/3}$), average degree $\av{k}$ (which
      determines $p_c = 1/\av{k}$), and disorder strength $a$ -- see
      Table~\ref{table:TableER} for details. Results were averaged
      over $1500$ realizations.}
  \end{figure}
}

\def\TableER{
  \begin{table}
    \begin{tabular}{c|c|c|c|c|c|c}
      $N$       &  $\av{k}$     &  $\ell_{\infty}$     &      $p_c$    &  $a$  &  $Z = \frac{1}{p_c}\frac{\ell_{\infty}}{a}$  &  Symbol  \\
      \hline
      \hline
      4000      &       3       &       42.48       &       1/3     & 12.73        &       10      &  x                         \\
      8000      &       3       &       60.59       &       1/3     & 18.16        &       10      &  $\Box$                    \\
      4000      &       5       &       44.01       &       1/5     & 22.00        &       10      &  $\bigtriangleup$          \\
      8000      &       5       &       58.42       &       1/5     & 29.19        &       10      &  $\ast$                    \\
      4000      &       8       &       45.99       &       1/8     & 36.78        &       10      &  $\diamond$                \\
      8000      &       8       &       58.25       &       1/8     & 46.60        &       10      &  $\circ$                   \\
      \hline
      4000      &       3       &       42.48       &       1/3     & 42.45        &       3       &  x                         \\
      8000      &       3       &       60.59       &       1/3     & 60.55        &       3       &  $\Box$                    \\
      4000      &       5       &       44.01       &       1/5     & 73.33        &       3       &  $\bigtriangleup$          \\
      8000      &       5       &       58.42       &       1/5     & 97.31        &       3       &  $\ast$                    \\
      2000      &       8       &       34.94       &       1/8     & 93.15        &       3       &  $\diamond$                \\
      4000      &       8       &       45.99       &       1/8     & 122.62       &       3       &  $\circ$                   \\
    \end{tabular}
    \caption{\label{table:TableER} Different disordered ER graphs with same
      value of $Z = \frac{1}{p_c}\frac{\ell_{\infty}}{a}$. The symbols refer
      to Fig.~\ref{fig:ERWeak}.}
  \end{table}
}

\def\figureSFWeak{
  \begin{figure}
    \resizebox{\subfigsize}{!}{\includegraphics{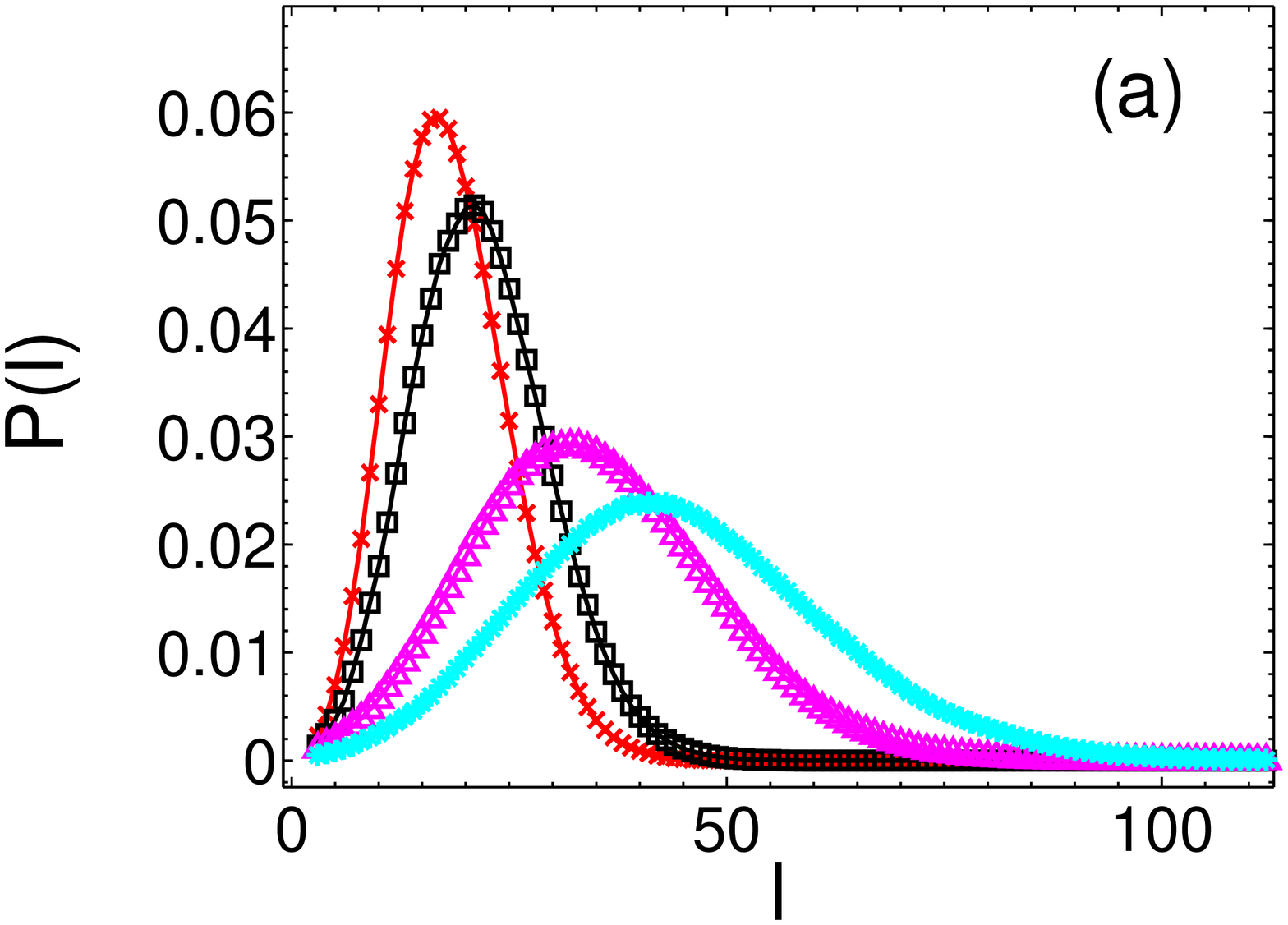}}
    \hskip 1 mm
    \resizebox{\subfigsize}{!}{\includegraphics{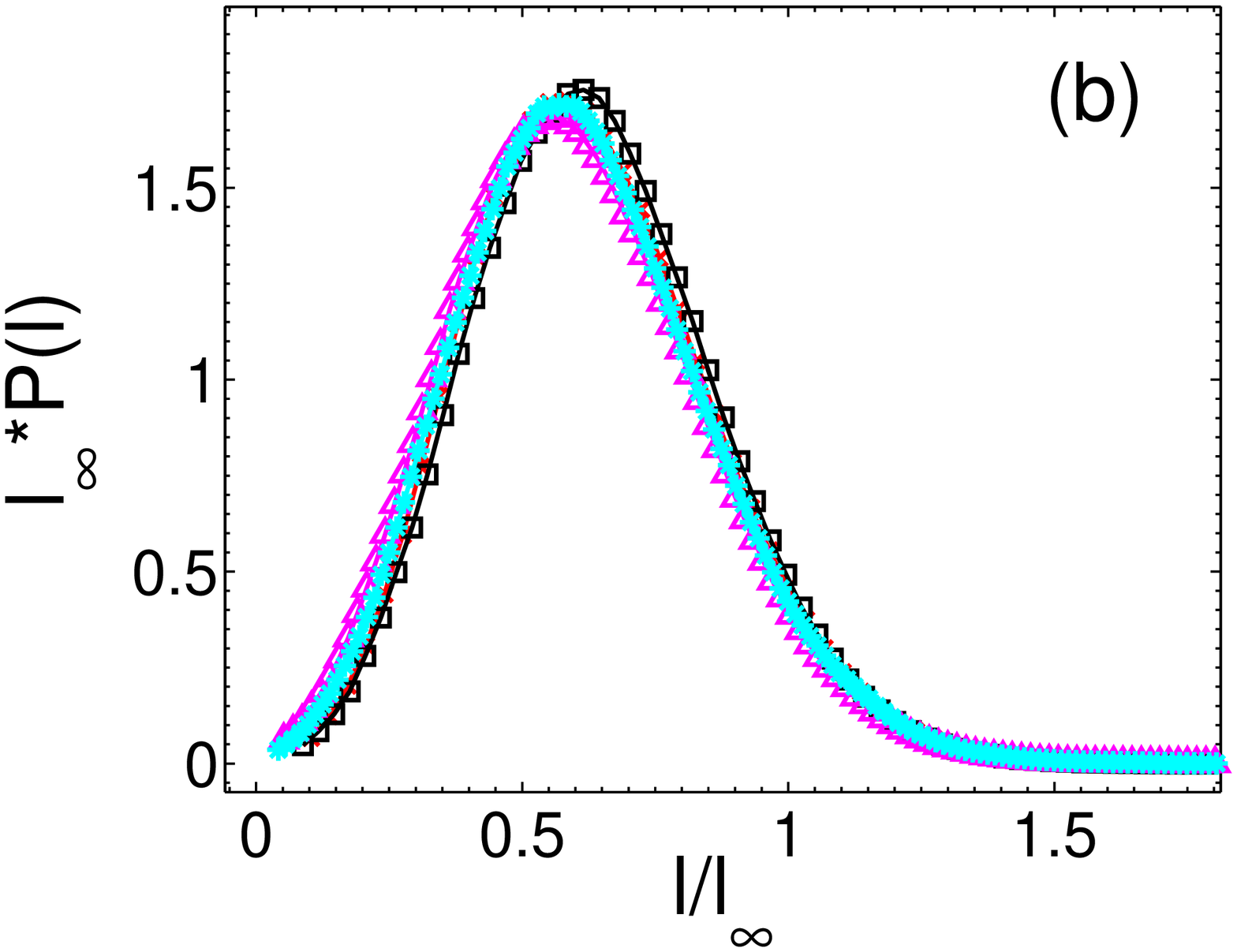}}
    \\
    \resizebox{\subfigsize}{!}{\includegraphics{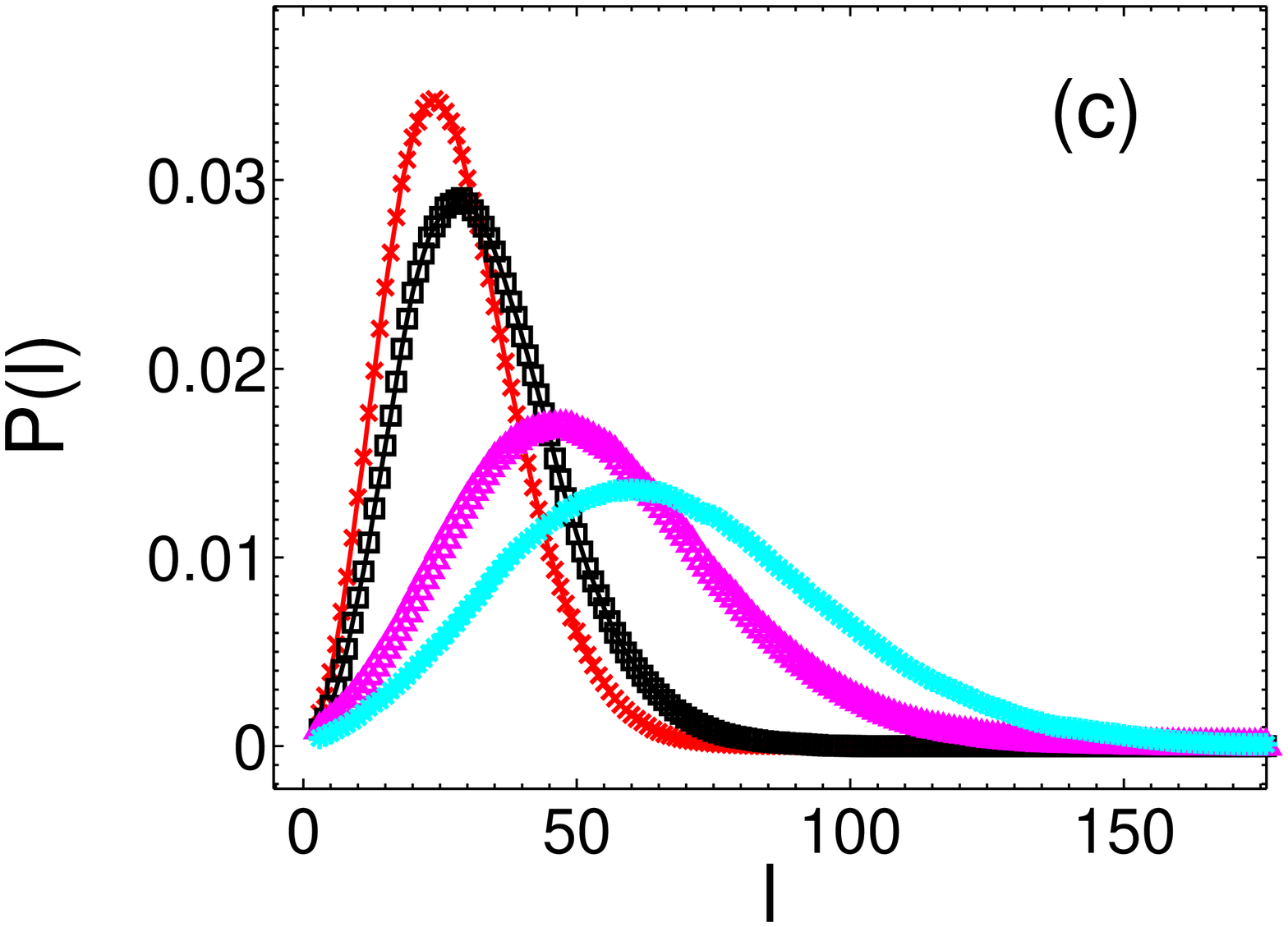}}
    \hskip 1 mm
    \resizebox{\subfigsize}{!}{\includegraphics{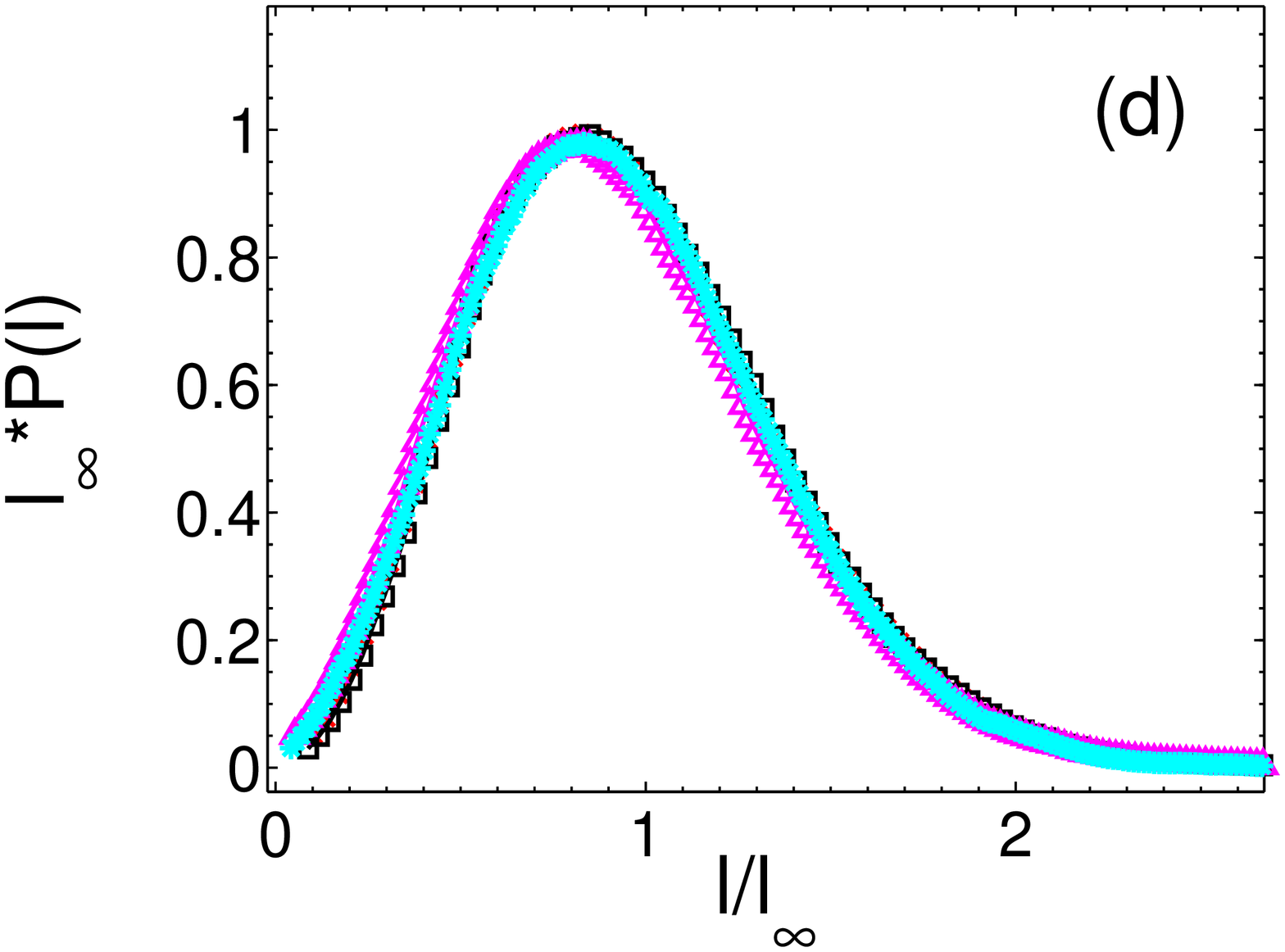}}
    \caption{\label{fig:SFWeak} (Color online) Optimal path lengths
      distribution, $P(l)$, for SF networks with (a,b) $Z \equiv
      \frac{1}{p_c}\frac{\ell_{\infty}}{a} = 10$ and (c,d) $Z =2$. (a)
      and (c) represent the unscaled distributions for $Z=10$ and
      $Z=2$ respectively, while (b) and (d) are the scaled
      distribution. Different symbols represent networks with
      different characteristics such as size $N$ (which determines
      $\ell_{\infty} \sim N^{\nu_{opt}}$), $\lambda$ and $m$ (which
      determine $p_c$), and disorder strength $a$ -- see
      Table~\ref{table:TableSF}. Results were averaged over $250$
      realizations.}
  \end{figure}
}

\def\TableSF{
  \begin{table}
    \begin{tabular}{c|c|c|c|c|c|c|c}
      $N$    &  $\lambda$  &  $m$ &  $\ell_{\infty}$  &   $p_c$    &   $a$    &  $Z = \frac{1}{p_c}\frac{\ell_{\infty}}{a}$  &  Symbol  \\
      \hline
      \hline
      4000   &     3.5     &  2   &    29.02       &    0.27    &  10.51   &       10      &  x                         \\
      8000   &     3.5     &  2   &    34.13       &    0.26    &  12.88   &       10      &  $\Box$                    \\
      4000   &     5       &  2   &    57.70       &    0.5     &  11.54   &       10      &  $\bigtriangleup$          \\
      8000   &     5       &  2   &    72.03       &    0.5     &  14.40   &       10      &  $\ast$                    \\
      \hline
      4000   &     3.5     &  2   &    29.02       &    0.27    &  52.56   &       2       &  x                         \\
      8000   &     3.5     &  2   &    34.13       &    0.26    &  64.44   &       2       &  $\Box$                    \\
      4000   &     5       &  2   &    57.70       &    0.5     &  57.70   &       2       &  $\bigtriangleup$          \\
      8000   &     5       &  2   &    72.03       &    0.5     &  72.03   &       2       &  $\ast$                    \\
      \end{tabular}
    \caption{\label{table:TableSF} Different disordered SF graphs with same
      value of $Z = \frac{1}{p_c}\frac{\ell_{\infty}}{a}$. The percolation threshold
      was calculated according to: $p_c = \frac{\av{k}}{\av{k(k-1)}}$. The symbols
      refer to Fig.~\ref{fig:SFWeak}.}
  \end{table}
}

\def\figureSFTwoFiveWeak{
  \begin{figure}
    \resizebox{\subfigsize}{!}{\includegraphics{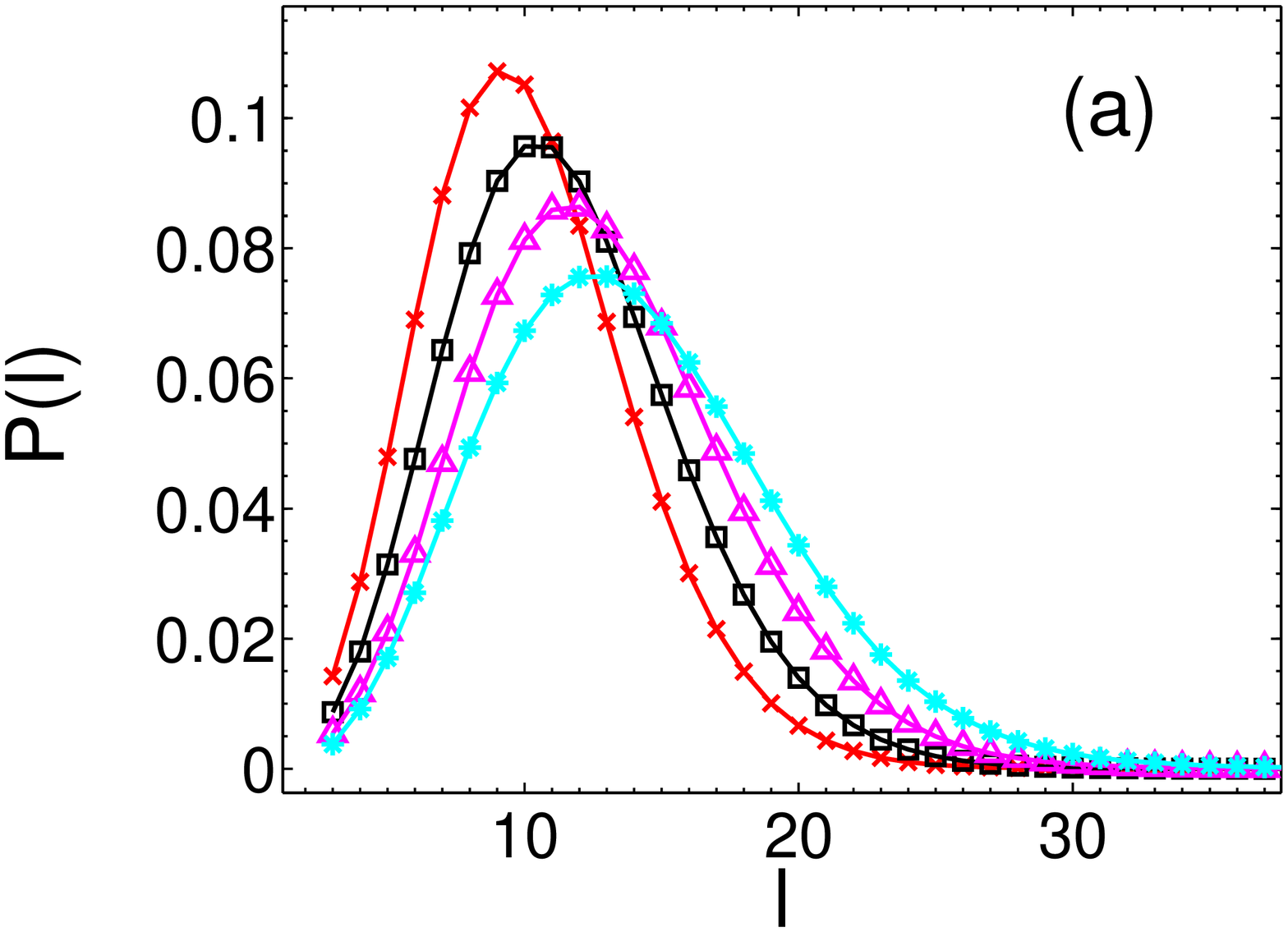}}
    \hskip 1 mm
    \resizebox{\subfigsize}{!}{\includegraphics{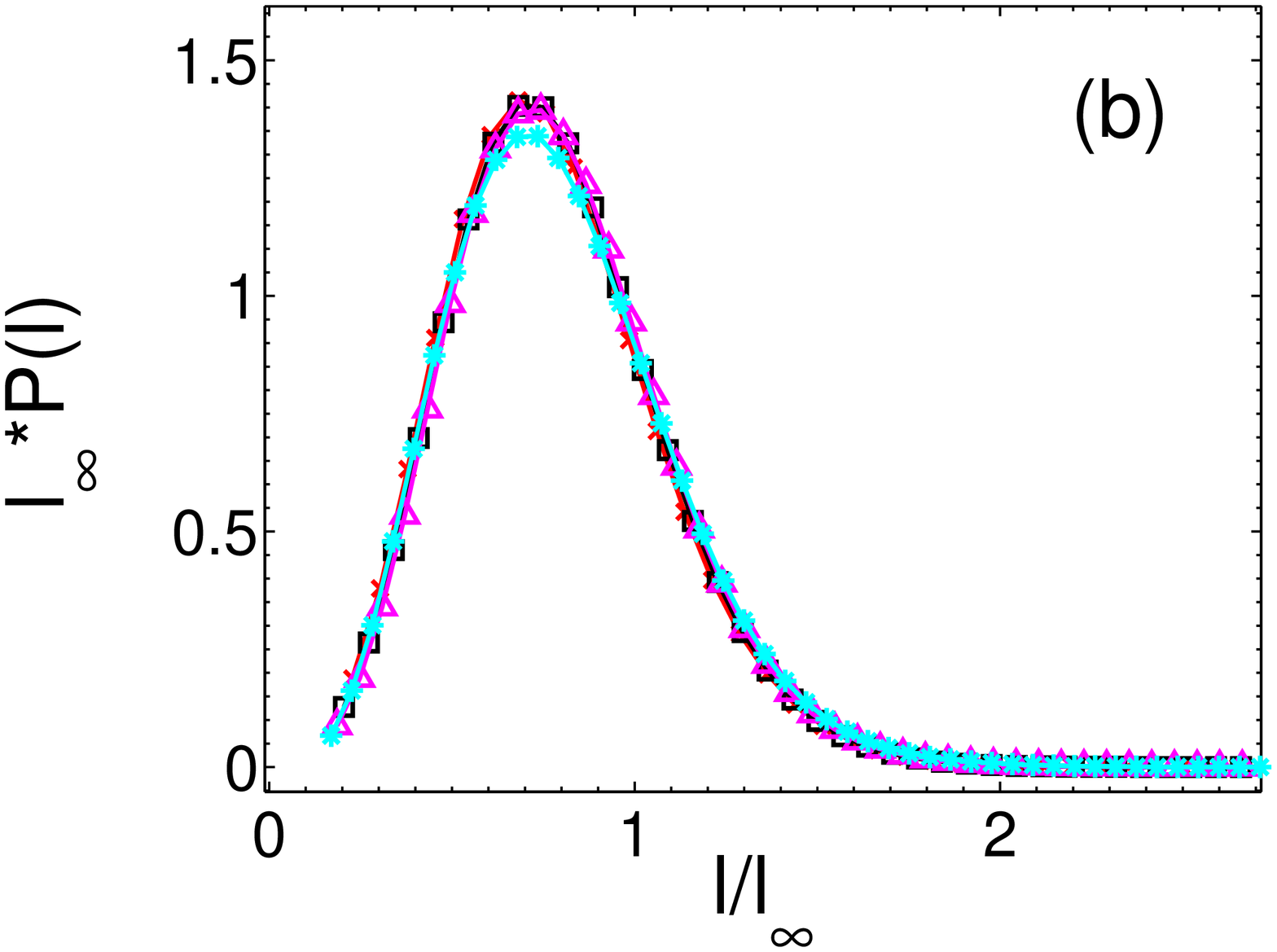}}
    \caption{\label{fig:SF25Weak} (Color online) Optimal path lenghts
      distribution function for SF graphs with $\lambda=2.5$, and with
      $Z \equiv \frac{1}{p_c}\frac{\ell_{\infty}}{a} = 10$. (a)
      represents the unscaled distribution for $Z=10$ while (b) shows
      the scaled distribution. Different symbols represent graphs with
      different characteristics such as size $N$ (which determines
      $\ell_{\infty} \sim \log(N)$ and $p_c \sim N^{-1/3}$), and
      disorder strength $a$ -- see Table~\ref{table:TableSFTwoFive}.
      Results were averaged over $1500$ realizations.}
  \end{figure}
}

\def\TableSFTwoFive{
  \begin{table}
    \begin{tabular}{c|c|c|c|c|c|c|c}
      $N$    &  $\lambda$  &  $m$ &  $\ell_{\infty}$  &   $p_c$    &   $a$    &  $Z = \frac{1}{p_c}\frac{\ell_{\infty}}{a}$  &  Symbol  \\
      \hline
      \hline
      2000   &     2.5     &  2   &   13.19        &   0.048    &   27.01  &       10      &  x                         \\
      4000   &     2.5     &  2   &   14.66        &   0.037    &   38.70  &       10      &  $\Box$                    \\
      8000   &     2.5     &  2   &   16.14        &   0.029    &   54.50  &       10      &  $\bigtriangleup$          \\
      16000  &     2.5     &  2   &   17.69        &   0.022    &   77.48  &       10      &  $\ast$                    \\
    \end{tabular}
    \caption{\label{table:TableSFTwoFive} Different disordered SF graphs
      with $\lambda=2.5$ and with same value of
      $Z = \frac{1}{p_c}\frac{\ell_{\infty}}{a}$. Notice that $p_c \sim N^{-1/3} \rightarrow 0$
      for $N \rightarrow \infty$. The symbols refer to Fig.~\ref{fig:SF25Weak}.}
  \end{table}
}

\def\figureTransitionSameGraph{
  \begin{figure}
    \resizebox{\subfigsize}{!}{\includegraphics{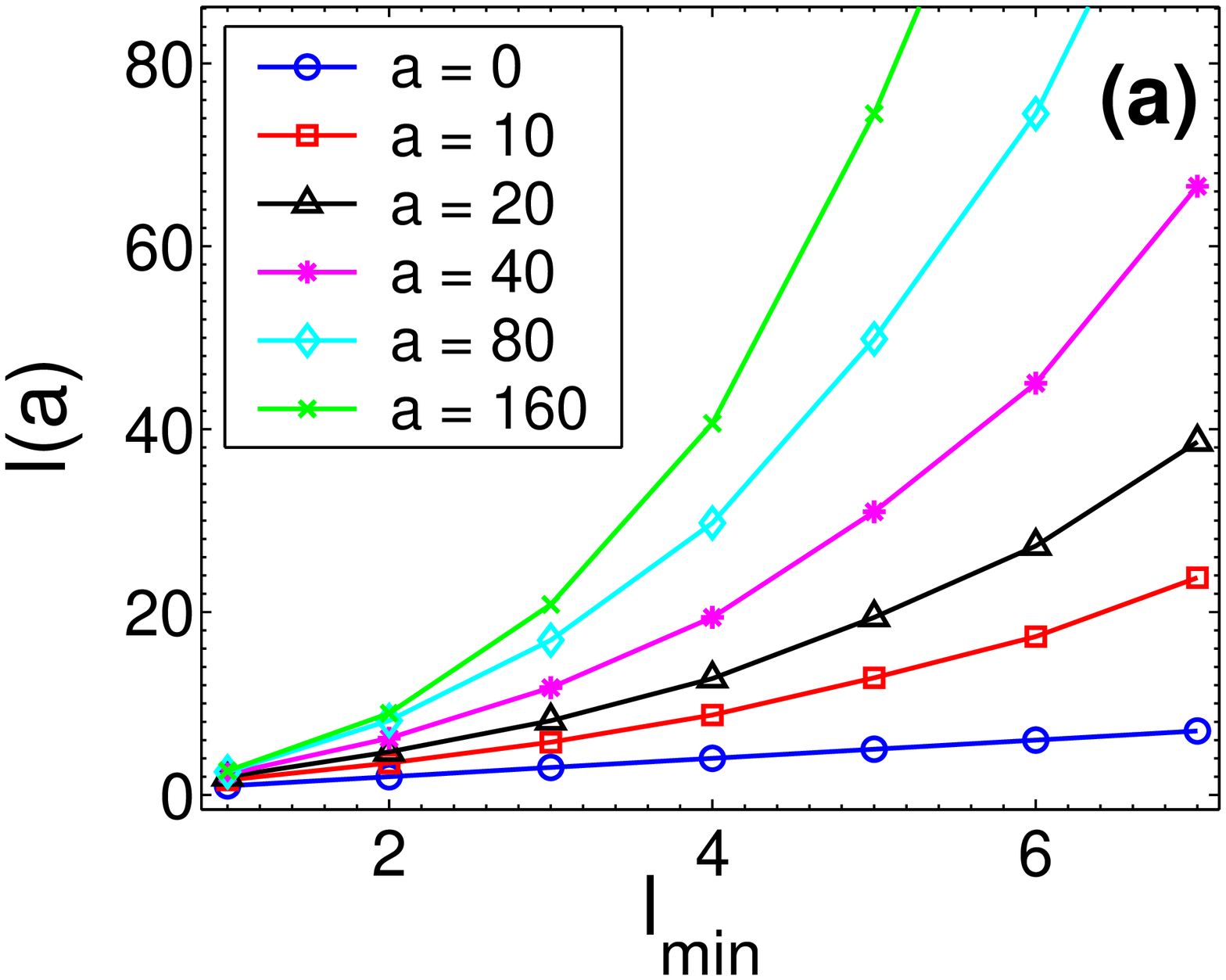}}
    \hskip 1 mm
    \resizebox{\subfigsize}{!}{\includegraphics{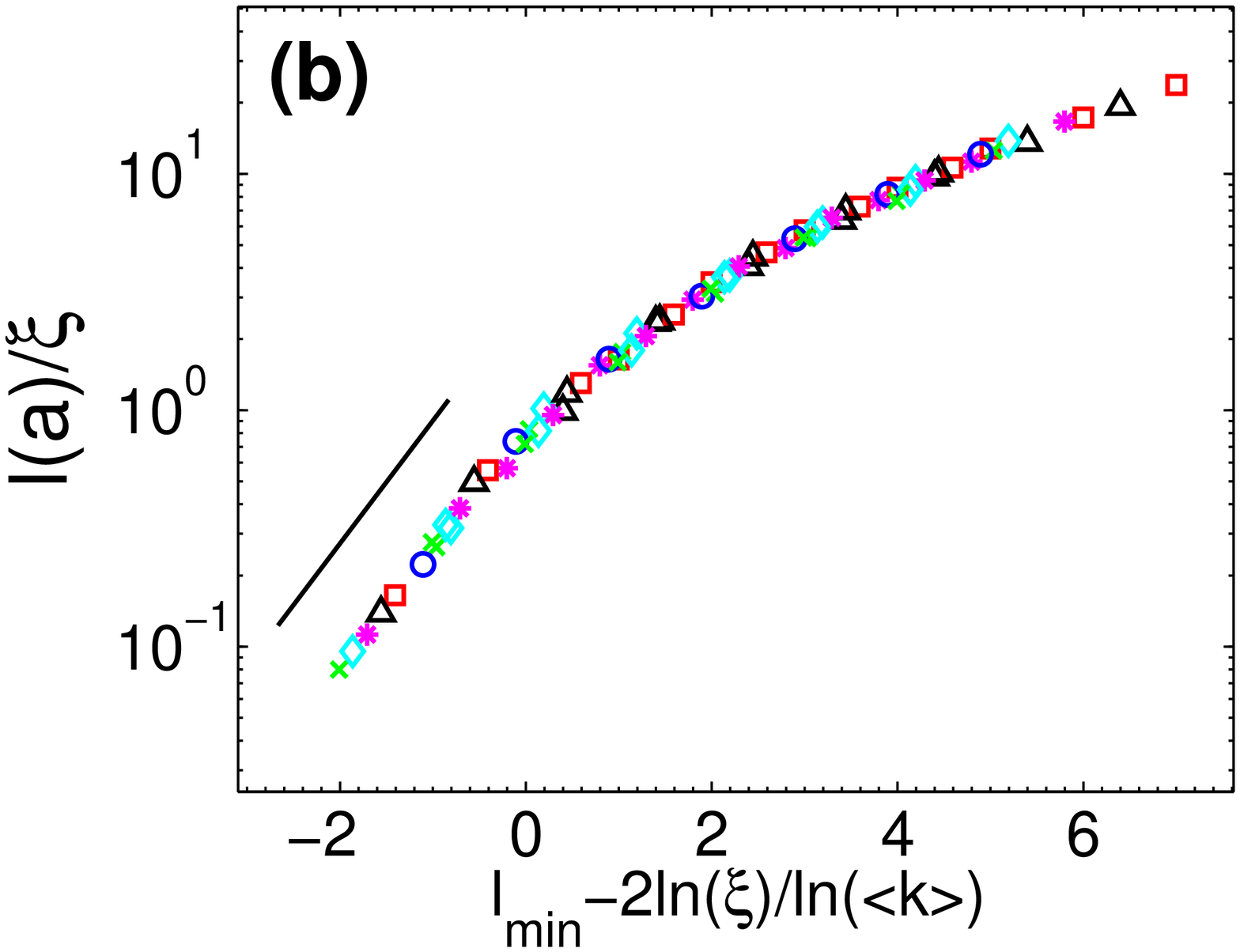}}
    \caption{\label{fig:TransitionSameGraph} (Color online) Transition
      between different scaling regimes for the optimal path length
      $l(a)$ inside an ER graph with $N=128,000$ nodes and
      $\av{k}=10$. (a) shows the unscaled and (b) shows the scaled
      length of the optimal path $l(a)$ averaged over all nodes with
      same value of $l_{min}$. Different symbols represent different
      values of the disorder strength $a$.
      Fig. (b) shows that for length scales $\ell(a)$ smaller than the
      ``characteristic length'', $\xi=ap_c$, $l(a)$ grows
      exponentially relative to the shortest hopcount path $l_{min}$
      (see solid line). This is consistent with $l(a) \sim N^{1/3}$
      and $l_{min} \sim \log{N}$ inside the range of size $\xi = a
      p_c$. For length scales above $\xi$ both quantities scale as
      $\log{N}$.  }
  \end{figure}
}

\section{\label{sec:Introduction}Introduction:}

Many real world systems exhibit a web-like structure and may be
treated as ``networks''. Examples may be found in physics, sociology,
biology, and
engineering~\cite{Barabasi-Albert-2002:statistical-mechanics,Dorogovtsev-Mendes-2003:From_Biological_Nets_to_the_Internet,vespignani-pastor-satorras-2004:evolution_and_structure}.
The function of most real world networks is to connect distant nodes,
either by transfer of information (e.g. the Internet), or through
transportation of people and goods (such as networks of roads and
airlines).  In many cases there is a ``cost'' or a ``weight''
associated with each link, and the larger the weight on a link, the
harder it is to traverse this link.  In this case, the network is
called ``disordered'' or
``weighted''~\cite{Braunstein-Buldyrev-Cohen-Havlin-Stanley-2003:Optimal,vespignani-2004:weighted_evolving_networks}.
For example, in the Internet each link between two routers has a
bandwidth or delay time, in a transportation network some roads may
have only one lane while others may be highways allowing for large
volumes of traffic.

The average length of the optimal path (or ``shortest path'') in
weighted lattices and networks has been extensively
studied~\cite{cieplak-maritan-banavar-1994:optimal_paths,cieplak-maritan-banavar-1996:invasion_percolation,porto-1997:optimal_path_strong_disorder,Braunstein-Buldyrev-Cohen-Havlin-Stanley-2003:Optimal,Sreenivasan-Kalisky-Braunstein-Buldyrev-Havlin-Stanley-2003:Effect}.
In weighted networks it is commonly assumed that each link is
associated with a weight $\tau_i = \exp(ar_i)$, where $r_i$ is a
random number taken from a uniform distribution between 0 and 1, and
the parameter $a$ controls the strength of the disorder. It has been
shown~\cite{Sreenivasan-Kalisky-Braunstein-Buldyrev-Havlin-Stanley-2003:Effect}
that the length of the optimal path in such weighted networks scales
as $l(a) \sim N^{\nu_{opt}}$ (where $\nu_{opt}$ is universal exponent)
for small system size $N$, and $l(a) \sim \log{N}$ for large
systems~\footnote{Throughout this paper, in cases where one quantity
  is proportional to the logarithm of another we will not specify the
  base of the logarithm explicitly, because the base may be changed
  arbitrarily by adjusting the constant of proportionality.}.  More
precisely:
\begin{equation}
  \label{equ:weighted_length}
  \ell(a) \sim \ell_{\infty} F\left( \frac{\ell_{\infty}}{ap_c} \right),
\end{equation}
where $p_c$ is the percolation threshold and $\ell_{\infty} \sim
N^{\nu_{opt}}$ is the optimal path length for strong disorder ($a
\rightarrow \infty$). For Erd\H{o}s-R\'enyi (ER) graphs
$\nu_{opt}=1/3$.  For scale-free (SF) networks, with a power law
degree distribution $P(k) \sim k^{-\lambda}$, $\nu_{opt} = (\lambda -
3)/(\lambda - 1)$ for $3<\lambda<4$ and $\nu_{opt} = 1/3$ for $\lambda
\geq 4$~\cite{Braunstein-Buldyrev-Cohen-Havlin-Stanley-2003:Optimal}.
The function $F(u)$ is of the form:
\begin{equation}
  \label{equ:small_word_f}
  F(u)=\left\{
    \begin{array}{l c c}
      \mathrm{const}      & \ \ \mathrm{if}\ \  &  u \ll 1         \\
      \log(u)/u           & \ \ \mathrm{if}\ \  &  u \gg 1
    \end{array} \right.
  .
\end{equation}

In this paper we study the following question: how are the different
optimal paths in a network distributed? The distribution of the
optimal path lengths is especially important in communication
networks, in which the overall network performance depends on the
different path lengths between all nodes of the network, and not only
the average. A recent work has studied the distribution form of
shortest path lengths on minimum spanning
trees~\cite{Braunstein-Buldyrev-Sreenivasan-Cohen-Havlin-Stanley-2004:the_optimal},
which correspond to optimal paths on networks with large variation in
link weights ($a \rightarrow \infty$).

We generalize these results and suggest that the distribution of the
optimal path lengths has the following scaling form:
\begin{equation}
  \label{equ:weighted_dist}
  P(\ell,N,a) \sim \frac{1}{\ell_{\infty}} G \left( \frac{\ell}{\ell_{\infty}} , \frac{1}{p_c}\frac{\ell_{\infty}}{a} \right).
\end{equation}
The parameter $Z \equiv \frac{1}{p_c}\frac{\ell_{\infty}}{a}$
determines the functional form of the distribution.
Relation~(\ref{equ:weighted_dist}) is supported by simulations for
both ER and SF graphs, including SF graphs with $2<\lambda<3$, for
which $p_c \rightarrow 0$ with system size
$N$~\cite{Cohen-Erez-Ben_Avraham-Havlin-2000:Resilience}
(Section~\ref{sec:simulations_probabilities}).

The paper is organized as follows: in
Section~\ref{sec:simulations_probabilities} we show results from
simulations for various ER and SF graphs. In
Section~\ref{sec:discussion} we explain these results and also show
that the optimal path $l_{opt}(a)$ \textit{inside a single network}
scales differently below and above a characteristic length $\xi=ap_c$.
For $\ell < \xi$ it is like strong disorder, while for $\ell > \xi$
the behavior is like weak disorder.

\section{\label{sec:simulations_probabilities}Erd\H{o}s-R\'enyi and Scale-Free Graphs:}

We simulate ER graphs with weights on the links for different values
of graph size $N$, control parameter $a$, and average degree $\av{k}$
(which determines $p_c=1/\av{k}$) -- see Table~\ref{table:TableER}. We
then generate the shortest path tree (SPT) using Dijkstra's
algorithm~\cite{Cormen-2001:Introduction} from some randomly chosen
root node. Next, we calculate the probability distribution function of
the shortest (i.e. optimal) path lengths for all nodes in the graph.

In Fig.~\ref{fig:ERWeak} we plot $\ell_{\infty}P(\ell,N,a)$ vs.
$\ell/\ell_{\infty}$ for different values of $N$, $a$, and $\av{k}$. A
collapse of the curves is seen for all graphs with the same value of
$Z = \frac{1}{p_c}\frac{\ell_{\infty}}{a}$.

\ifnum\tipo=2
\figureERWeak
\fi

\ifnum\tipo=2
\TableER
\fi

Figure~\ref{fig:SFWeak} shows similar plots for SF graphs -- with a
degree distribution of the form $P(k) \sim k^{-\lambda}$ and with a
minimal degree $m$~\footnote{Scale-free graphs were generated
  according to the ``configuration model''
  (e.g.~\cite{Bollobas-1980:model,Molloy-Reed-1998:Size,Sokolov-2003:Tomography,Kalisky-Cohen-ben-Avraham-Havlin-2004:Tomography}).
  In this method, each node is assigned a number of open ``stubs''
  according to the scale-free degree distribution $P(k)$. Then, these
  stubs are interconnected randomly, thus creating a network having
  the required degree distribution $P(k)$.}~\footnote{Note that the
  minimal degree is $m=2$ thus ensuring that there exists an infinite
  cluster for any $\lambda$, and thus $0<p_c<1$. For the case of $m=1$
  there is almost surely no infinite cluster for $\lambda > \lambda_c
  \approx 4$ (or for a slightly different model, $\lambda_c =
  3.47875$~\cite{aiello-chung-lu-2000:random_graph_model}), resulting
  in an effective percolation threshold $p_c =
  \frac{\av{k}}{\av{k(k-1)}} > 1$.
  See~\cite{Kalisky-Cohen-ben-Avraham-Havlin-2004:Tomography,aiello-chung-lu-2000:random_graph_model}
  for details.}. A collapse is obtained for different values of $N$,
$a$, $\lambda$ and $m$, with $\lambda>3$ (see
Table~\ref{table:TableSF}).

\ifnum\tipo=2
\figureSFWeak
\fi

\ifnum\tipo=2
\TableSF
\fi

Next, we study SF networks with $2 < \lambda < 3$. In this regime the
second moment of the degree distribution $\langle k^2 \rangle$
diverges, leading to several anomalous
properties~\cite{Cohen-Erez-Ben_Avraham-Havlin-2000:Resilience,Cohen-Havlin-2003:Ultra,newman-callaway-2000:networks_robustness}.
For example: the percolation threshold approaches zero with system
size: $p_c \sim N^{-\frac{3-\lambda}{\lambda-1}} \rightarrow 0$, and
the optimal path length $\ell_{\infty}$ was found numerically to scale
logarithmically (rather than polynomially) with
$N$~\cite{Braunstein-Buldyrev-Cohen-Havlin-Stanley-2003:Optimal}.
Nevertheless, as can be seen from Fig.~\ref{fig:SF25Weak} and
Table~\ref{table:TableSFTwoFive}, the optimal paths probability
distribution for SF networks with $2 < \lambda < 3$ exhibits the same
collapse for different values of $N$ and $a$ (although its functional
form is different than for $\lambda>3$).

\ifnum\tipo=2
\figureSFTwoFiveWeak
\fi

\ifnum\tipo=2
\TableSFTwoFive
\fi

\section{\label{sec:discussion}Discussion:}

We present evidence that the optimal path is related to
percolation~\cite{Sreenivasan-Kalisky-Braunstein-Buldyrev-Havlin-Stanley-2003:Effect}.
Our present numerical results suggest that for a finite disorder
parameter $a$, the optimal path (on average) follows the percolation
cluster in the network (i.e., links with weight below $p_c$) up to a
typical ``characteristic length'' $\xi = a p_c$, before deviating and
making a ``shortcut'' (i.e. crossing a link with weight above $p_c$).
For length scales below $\xi$ the optimal path behaves as in strong disorder
and its length is relatively long. The shortcuts have an effect of
shortening the optimal path length from a polynomial to logarithmic
form according to the universal function $F(u)$
(Eq.~\ref{equ:small_word_f}).
\omitit{~\footnote{This is somewhat
    analogous to the small world model, see
    e.g.~\cite{barthelmy-amaral-1999:small_world_crossover}.}}
Thus, the optimal path for finite $a$ can be viewed as consisting of
``blobs'' of size $\xi$ in which strong disorder persists. These blobs
are interconnected by shortcuts, which result in the total path being
in weak disorder.

We next present direct simulations supporting this argument. We
calculate the optimal path length $l(a)$ inside a single network, for
a given $a$, and find (Fig.~\ref{fig:TransitionSameGraph}) that it
scales differently below and above the characteristic length $\xi =
ap_c$.  For each node in the graph we find $l_{min}$, which is the
number of links (``hopcounts'') along the shortest path from the root
to this node \textit{without regarding the weight of the
  link}
~\footnote{This is done by using the Breadth-First-Search (BFS)
  algorithm~\cite{Cormen-2001:Introduction}.}
. In Fig.~\ref{fig:TransitionSameGraph} we plot the length of the
optimal path $l(a)$, averaged over all nodes with the same value of
$l_{min}$ for different values of $a$. The figure strongly suggests
that $l(a) \sim \exp(l_{min})$ for length scales below the
characteristic length $\xi = a p_c$, while for large length scales
$l(a) \sim l_{min}$
\footnote{\label{foot:transition_same_graph} For length scales smaller
  than $\xi$ we have $l_{opt}=A N^{1/3}$ and $l_{min} = B \ln{N}$,
  where $A$ and $B$ are constants.  Thus $N = \exp{(l_{min}/B)}$ and
  $l_{opt} = A \exp{(l_{min}/3B)}$.  Consequently, we expect that:
  $\frac{l_{opt}}{\xi} = \frac{A \exp{(l_{min}/3B)}}{\xi} = A
  \exp{[(l_{min}-3B \ln{\xi})/3B]}$.  We find the best scaling in
  Fig.~\ref{fig:TransitionSameGraph} for $B =
  \frac{2}{3\ln{\av{k}}}$.}
.
This is consistent with our hypothesis that below the characteristic
length ($\xi = a p_c$) $l_{min} \sim \log{N}$ and $l(a) \sim N^{1/3}$,
while $l_{min} \sim \log{N}$ and $l(a) \sim \log{N}$ above.

\ifnum\tipo=2
\figureTransitionSameGraph
\fi

In order to better understand why the distributions of $l_{opt}$
depend on $Z$ according to Eq.~(\ref{equ:weighted_dist}), we suggest
the following argument.  The optimal path for $a \rightarrow \infty$,
was shown to be proportional to $N^{1/3}$ for ER graphs and
$N^{(\lambda - 3)/(\lambda - 1)}$ for SF graphs with
$3<\lambda<4$~\cite{Braunstein-Buldyrev-Cohen-Havlin-Stanley-2003:Optimal}.
For finite $a$ the number of shortcuts, or number of blobs, is $Z =
\frac{\ell_{\infty}}{\xi} = \frac{\ell_{\infty}}{ap_c}$.  The
deviation of the optimal path length for finite $a$ from the case of
$a \rightarrow \infty$ is a function of the number of shortcuts.
These results explain why the parameter $Z \equiv
\frac{\ell_{\infty}}{ap_c}$ determines the functional form of the
distribution function of the optimal paths.

\section{\label{sec:summary}Summary and Conclusions:}

To summarize, we have shown that the optimal path length distribution
in weighted random graphs has a universal scaling form according to
Eq.~(\ref{equ:weighted_dist}). We explain this behavior and
demonstrate the transition between polynomial to logarithmic behavior
of the average optimal path in a single graph.
Our results are consistent with results found for finite dimensional
systems~\cite{porto-1999:optimal_path_order_disorder,wu-lopez-2005:current_flow,strelniker-havlin-berkovits-frydman-2005:Tomography,perlsman-havlin-2005:directed_polymer}:
In finite dimension the parameter controlling the transition is
$\frac{L^{1/\nu}}{ap_c}$, where $L$ is the system length and $\nu$ is
the correlation length critical exponent (for random graphs $\nu=1$
when calculated in the shortest path metric). This is because only the
``red bonds'' - bonds that if cut would disconnect the percolation
cluster~\cite{coniglio-1982:cluster_structure} - control the
transition.

\section*{Acknowledgments}
We thank the ONR, the Israel Science Foundation, and the Israeli
Center for Complexity Science for financial support. We thank R.
Cohen, S. Sreenivasan, E.~Perlsman and Y. Strelniker for useful
discussions. Lidia A.~Braunstein thanks the ONR - Global for financial
support. This work was also supported by the European research NEST
Project No. DYSONET 012911.

\appendix

\bibliography{distributions} 
  
  
  

\ifnum\tipo=1
\figureERWeak
\TableER
\figureSFWeak
\TableSF
\figureSFTwoFiveWeak
\TableSFTwoFive
\figureTransitionSameGraph
\fi

\end{document}